\documentclass[preprint,12pt]{elsarticle}
\usepackage{amssymb,epsfig,graphicx}
\journal{Journal of Magnetism and Magnetic Materials}
\begin{document}
\begin{frontmatter}

\title{Effect of carbon content on magnetostructural properties of Mn$_3$GaC}
\author[gu]{E. T. Dias},
\author[gu]{K. R. Priolkar\corref{krp}}\ead{krp@unigoa.ac.in},
\author[tifr]{A. K. Nigam}
\cortext[krp]{Corresponding author}
\address[gu]{Department of Physics, Goa University, Goa 403206, India.}
\address[tifr]{Tata Institute of Fundamental Research, Dr. Homi Bhabha Road, Colaba, Mumbai 400005, India.}

\begin{abstract}
Effect of carbon content on magnetostructural transformation in antiperovskites of the type Mn$_3$GaC$_x$ ($x$ = 0.8, 1.0 and 1.05) has been investigated. It is found that, increase in carbon content changes the ground state from ferromagnetic metallic ($x$ = 0.8) to antiferromagnetic semiconducting ($x$ = 1.05) type. This has been attributed to localization of itinerant Mn $3d$ electrons due to increased Mn$3d$ - C$2p$ hybridization. Such a hybridization strengthens Mn-C-Mn antiferromagnetic interactions over Mn-Mn ferromagnetic interactions. Further, magnetic field can be used as a tool to modulate the relative strengths of these ferromagnetic and antiferromagnetic interactions thereby affecting the nature and strength of magnetocaloric properties.
\end{abstract}

\end{frontmatter}

\section{Introduction}
The antiperovskite Mn$_{3}$GaC has a cubic structure \cite{e-dias1,e-dias10} and belongs to a class of materials that are gaining interest due to their magnetocaloric properties \cite{e-dias5,e-dias6,e-dias7,e-dias8,e-dias11}. Stoichiometric composition of Mn$_3$GaC undergoes a series of magnetic transformations as a function of temperature. Firstly, it undergoes a second order transformation from a paramagnetic (PM) state to a ferromagnetic (FM) state at $T_{C}\sim 249K$ followed by another second order transformation to a canted ferromagnetic state (CFM) at $T\sim 164K$. Finally the material undergoes a volume discontinuous first order magnetic transformation to an  an antiferromagnetic (AFM) state at $T\sim 160K$ \cite{e-dias3,e-dias4}. Concentration of carbon plays a vital role in inducing the first order magnetic transformation. Literature suggests that samples deficient in carbon, Mn$_3$GaC$_{0.8}$, exhibits only a second order PM to FM transition at $T_C\sim 318K$ \cite{e-dias5} while the compound with carbon concentration equal to 1.0, undergoes additional magnetic transitions leading to an antiferromagnetic ground state. The first-order magnetic transition from a ferromagnetic to antiferromagnetic state in Mn$_3$GaC produces a large change in entropy in presence of magnetic field giving rise to a large magnetocaloric effect (MCE), a property useful in magnetic refrigeration \cite{e-dias6,e-dias7}.

Although the role of carbon content in inducing the magnetic phase transformation has been contemplated in literature \cite{e-dias1,e-dias12}, the exact nature of the driving force in inducing the magnetostructural transformation is not yet understood. Neutron diffraction studies have shown that increase in C concentration results in a lattice expansion, increasing Mn-Mn next nearest neighbor distance. This is implied to cause an increase in Mn $3d$ - C $2p$ hybridization which strengthens Mn-C-Mn antiferromagnetic interactions over Mn-Mn ferromagnetic interactions causing the sample to transform via a discontinuous volume expansion \cite{e-dias1,e-dias7,e-dias13,e-dias14,e-dias16}. However, X-ray absorption near edge structure (XANES) studies at Mn L edges did not show any change in near edge features as a function of C content that would suggest an increase in Mn $3d$ - C $2p$ hybridization \cite{e-dias5}. It must be mentioned here that these studies were carried out on compositions wherein carbon content varied from $0.8 \le x \le 1.0$. In this paper for the first time we examine Mn$_3$GaC$_x$ with $x$ = 1.05 (C-excess) along with $x$ = 0.8 (C-deficient) and $x$ = 1.0 (C-stoichiometric) compounds. From structural, transport and magnetic properties of these three compositions, it is shown that increase in carbon concentration leads to an increase in cell volume, a change in character of resistivity from metallic in $x$ = 0.8 to semiconducting in $x$ = 1.05 and dominance of antiferromagnetic interactions. These characteristics point towards localization of conduction electrons and highlight the importance of Mn-C-Mn interactions in inducing the first order magnetic transformation from ferromagnetic to antiferromagnetic state. Furthermore, it is also shown that magnetic field can be effectively used as a tool to modulate the nature of magnetic interactions and therefore the magnitude and nature of magnetocaloric properties.


\section{Experimental}

The polycrystalline samples of Mn$_3$GaC$_x$ were prepared by solid state reaction method. First the constituent elements, Mn, Ga and C were thoroughly mixed in molar ratio of 3:1:1, pressed into pellets, sealed in evacuated quartz tubes and heated for five days at 1073K. The sample was then quenched to room temperature, pulverised, pelletized and annealed again for five more days at 1073K \cite{e-dias7}. A comparison of results obtained from X-ray diffraction studies, density measurements and magnetization measurements (described later) with those reported in literature \cite{e-dias4} showed that the resulting compound was sub stoichiometric having a composition Mn$_3$GaC$_{0.8}$. After preserving a third of this powder, the other two parts were respectively mixed with additional carbon equal to 0.20 and 0.25 times the original carbon content  and were further annealed at 1073K for five days in evacuated and sealed quartz tubes to obtain Mn$_3$GaC and Mn$_3$GaC$_{1.05}$. X-ray diffraction (XRD) patterns, to identify the crystal structure and phase purity of the powdered samples were recorded at room temperature in the range of $ 20^\circ \leq 2\theta \leq 80^\circ$ using Cu K$_\alpha $ radiation and at Indian beamline, BL18B, Photon Factory, Japan at photon energy of 13 keV. In order to study the effect of carbon content on the transport properties, resistivity measurements (10K - 330K) on all the samples were performed using conventional D.C. four probe technique in a Closed Cycle Refrigerator (CCR). To investigate the role of carbon on magnetic properties of these samples, magnetization (M) measurements were carried out as a function of temperature (T) and magnetic field (H) in the temperature range of 5K to 300K and in fields up to 7T using a Quantum Design SQUID magnetometer and a Quantum Design Vibrating sample magnetometer up to fields of 14T.

\section{Results and Discussion}

\begin{figure}[h]
\centering
\includegraphics[width=\columnwidth]{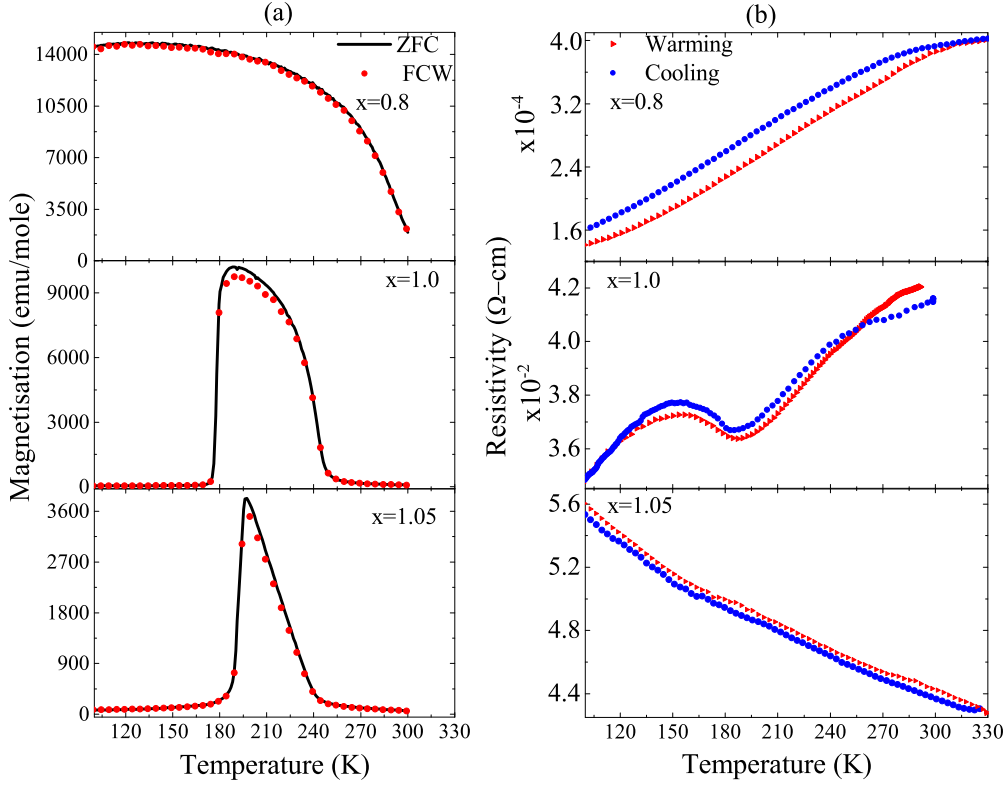}
\caption{\textbf{(a)} Temperature dependence of the magnetization M(T) during ZFC and FCW cycles in 0.1T for Mn$ _{3} $GaC$ _{x} $, $x$ = 0.8, 1.0 and 1.05. \textbf{(b)} Variation of electrical resistivity as a function of temperature for Mn$ _{3} $GaC$ _{x} $, $x$ = 0.8, 1.0 and 1.05.}
\label{fig:mt-rt}
\end{figure}

Figure \ref{fig:mt-rt} depicts temperature dependence of magnetization and resistivity of Mn$_3$GaC$_{x}$ ($x$ = 0.80, 1.0 and 1.05). Temperature dependent magnetization curves M(T) for the three compositions was measured in an applied field of 0.1T during zero-field cooling (ZFC) and field cooled warming (FCW) cycles (Figure \ref{fig:mt-rt}(a)). The C-deficient compound shows a PM to FM transition with $T_{C}$ just above 300K, whereas the C-stoichiometric compound shows the same PM to FM transition at a lower temperature ($ T_{C} = 242K $) followed by a FM to AFM first-order magnetic transition at $T_N = 178K$. Interestingly, in case of C-excess sample, a clear PM to FM transition is absent. It appears as if the sample transforms to an AFM ground state via a first-order transition at $T_N = 195K$ directly from a PM state. Such a behaviour has been observed in related antiperovskites Mn$_3$Cu$_{0.5}$Ge$_{0.5}$N and Mn$_3$GaC$_{0.85}$N$_{0.15}$ \cite{nitride,nitride1}. However, ferromagnetic correlations are present especially below 240K which manifest in terms of a relatively sharper rise in magnetization as can be seen in M versus T plot of Mn$_3$GaC$_{1.05}$ presented in Figure \ref{fig:mt-rt}(a). It may be noted that this temperature at which ferromagnetic correlations appear is lower than the T$_C$ of C-stoichiometric sample.   Thus it can be inferred that with increase in C content, the strength of FM interactions decreases while the strength of AFM interactions increases.

Figure \ref{fig:mt-rt}(b) compares the electrical resistivity behavior of the three compounds as a function of temperature. It can be seen that the magnitude of resistivity increases with increasing C content. The C-deficient sample has the lowest value of resistance and exhibits a metallic behavior with a distinct slope change around T$_C \sim$ 300K depicting PM to FM transition. Likewise, the resistivity of C-stoichiometric sample also has an overall positive temperature coefficient but with a magnitude which is about an order of magnitude higher than the C-deficient sample. This resistivity behavior is quite similar to that reported for Mn$_3$GaC composition in literature \cite{e-dias2}. Two features that stand out are a subtle change in slope around $ T_{C}$ ($\sim$ 240K) followed by a sharp increase in resisitance and a hysteresis between warming and cooling cycles near the first order transformation temperature. The higher resistance at lower temperature hints at localization of conduction electrons in the high volume cubic phase to which the compound transforms at T$_N$. On the other hand C-excess compound not only has the highest magnitude of resistivity but also displays a completely opposite variation of resistivity with temperature. Here the resistivity increases with decrease in temperature which is akin to semiconductor like behavior. A weak discontinuity is also visible around 190K corresponding to the first order transformation to antiferromagnetic state. The gradual increase in resistivity as a function of carbon content and the change in temperature gradient of resistivity from positive to negative with increase in carbon content is a signature of the localization of $3d$ conduction electrons in the carbon valence band.

\begin{figure}[h]
\centering
\includegraphics[width=\columnwidth]{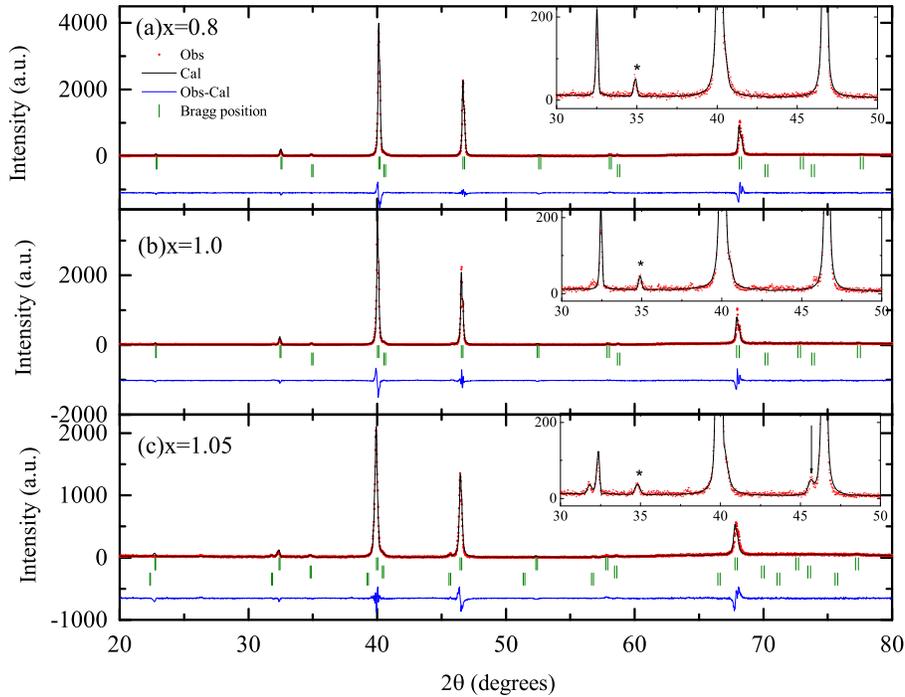}
\caption{Room temperature XRD patterns of Mn$ _{3} $GaC$ _{x} $, $x$ = 0.8, 1.0 and 1.05 refined using LeBail method. Insets show diffraction pattern for $ 30^\circ \leq 2\theta \leq 50^\circ$. Astericks indicate the MnO phase and the additional high volume phase in the C-excess sample is marked with an arrow.}
\label{fig:xrd}
\end{figure}
\begin{figure}[h]
\centering
\includegraphics[width=\columnwidth]{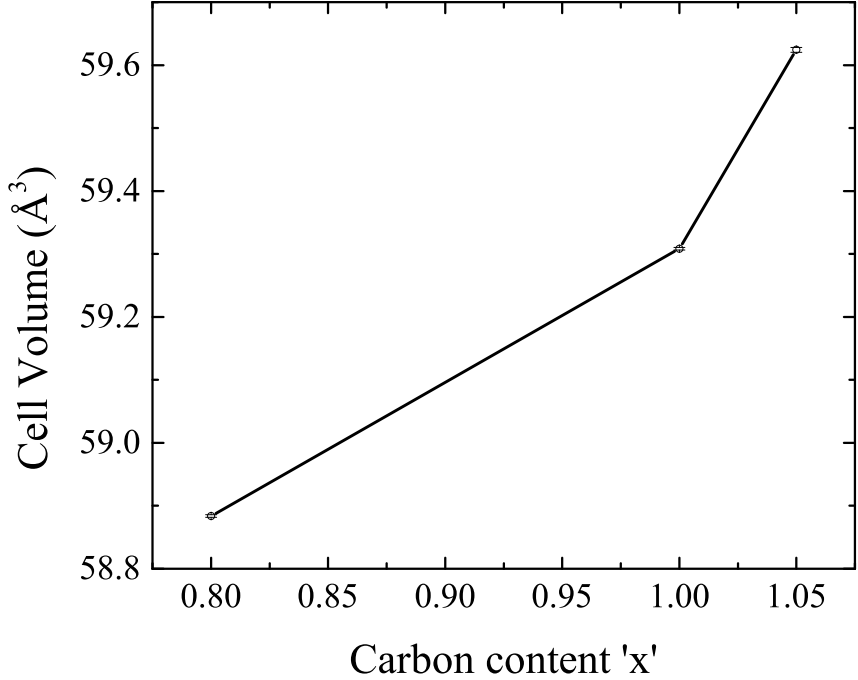}
\caption{Variation of unit cell volume with carbon concentration for Mn$ _{3} $GaC$ _{x} $, $x$ = 0.8, 1.0 and 1.05.} \label{fig:lattice}
\end{figure}

In order to understand the variations in magnetic and transport properties exhibited by these compounds room temperature XRD patterns of Mn$ _{3} $GaC$ _{x} $ ($x$ = 0.80, 1.0 and 1.05) were refined using the LeBail method in FullProf Suite. It can be seen in Figure \ref{fig:xrd} that all the three compositions crystallize largely in a simple cubic Mn$ _{3} $GaC phase with space group \textit{Pm3m} \cite{e-dias1} along with a tiny impurity ($<$ 1\%) of MnO. Such an insignificant amount of impurity does not to affect the magnetic and transport properties as can be seen in Figure \ref{fig:mt-rt} and which are in good agreement with those reported in literature \cite{e-dias5,e-dias2}. Although very little changes in the intensities of Bragg peaks were observed for compounds with varying carbon concentration, the unit cell volume and the density calculated from XRD exhibit an increasing trend with increase in carbon content. While the density increases from 6.95 g/cm$^3$ to 6.97 g/cm$^3$ with increase in carbon content, the variation of unit cell volume is depicted in Figure \ref{fig:lattice}. Increase in cell volume  will directly increase the Mn-Mn bond distance and cause the ferromagnetic interactions to weaken\cite{e-dias17}. Ferromagnetic interactions in these compounds is believed to be mediated by itinerant electrons \cite{e-dias2}. With increase in carbon content, the conduction electrons seem to get localized resulting in weakening of ferromagnetism and a concomittant increase in resistivity. Indeed such a trend is readily seen from the resistivity and magnetization measurements. Furthermore, a satisfactory LeBail fit for Mn$_3$GaC$_{1.05}$ could be obtained only by taking into account an additional antiperovskite cubic phase with a higher cell volume. The refined lattice parameter obtained for this phase were approximately equal to that of the transformed antiferromagnetic phase of Mn$_3$GaC \cite{e-dias1}. This high volume antiperovskite phase is antiferromagnetic in nature and its presence could be the reason for the absence of a clear ferromagnetic ordering in Mn$3$GaC$_{1.05}$.

Apart from the decrease in T$_C$, it is also observed that antiferromagnetic ordering not only appears but also increases in strength with increasing carbon content. The C-deficient compound can be conceived as consisting of randomly distributed vacancies in the carbon sublattice and relatively fuller Mn and Ga sublattices. Therefore the ferromagnetic Mn-Mn interactions dominate and there is no transformation to the antiferromagnetic state. While in case of the C-stoichiometric sample where all the individual sublattices are likely have near optimal occupation, a strong competition between ferromagnetic Mn-Mn and antiferromagnetic Mn-C-Mn interactions exists. In the C-excess sample, presence of additional carbon further strengthens antiferromagnetic interactions leading to an increase in T$_N$.
It is reported that due to increase in Mn-Mn bond distance the hybridization between Mn $3d$ and C $2p$ bands increases inducing antiferromagnetic Mn-C-Mn superexchange interactions which compete with ferromagnetic interactions. This competition drives the compound to undergo a  first order magnetic transformation from ferromagnetic cubic to an antiferromagnetic cubic phase of higher volume. Furthermore, the localization of itinerant Mn $3d$ conduction electrons due to Mn $3d$ - C $2p$ hybridization is clearly brought out from the resistivity measurements. Here, not only does the magnitude of resistivity increases with increasing carbon content, its character itself changes from metallic to one displaying semiconductor like behavior.


\begin{figure}[h]
\centering
\includegraphics[width=\columnwidth]{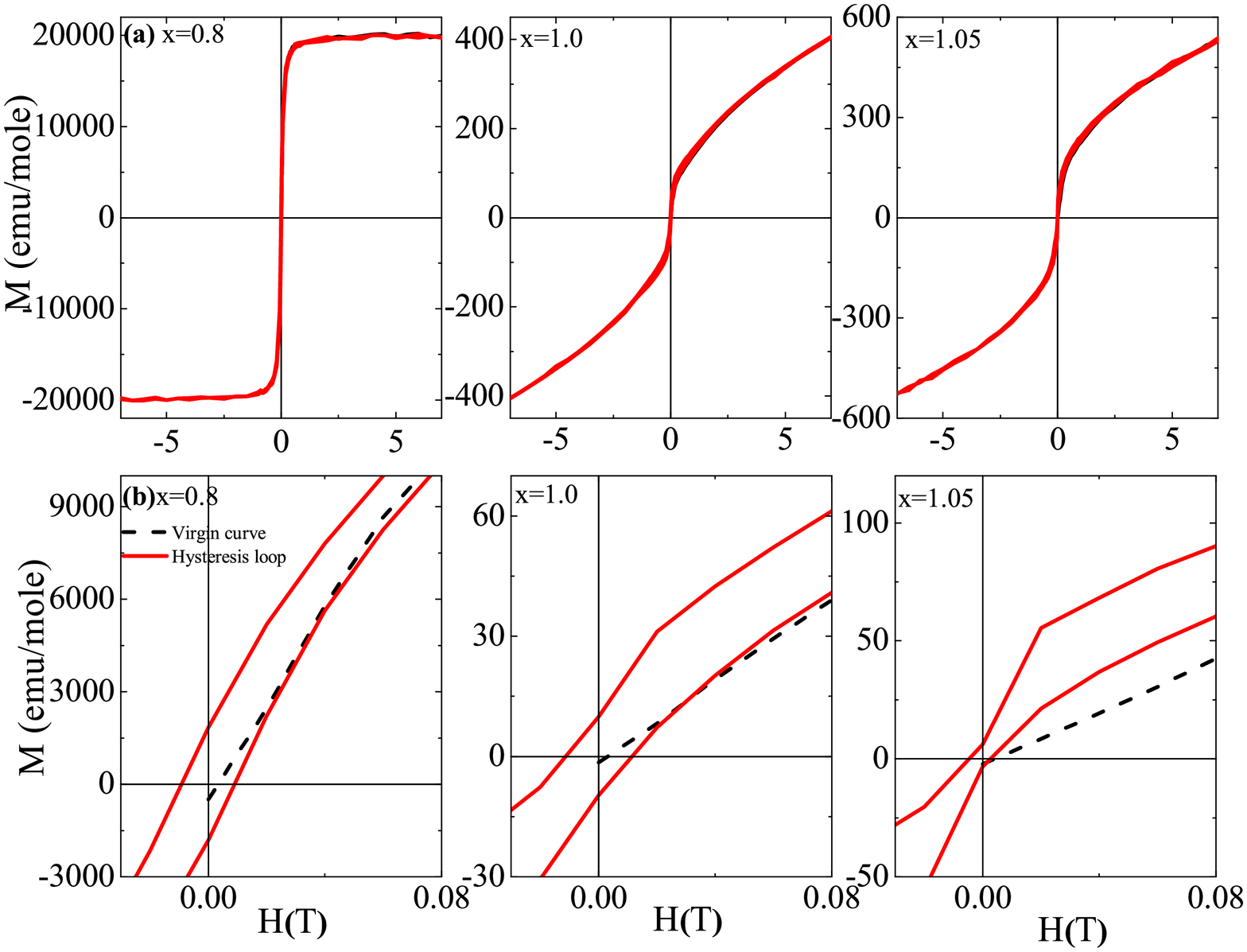}
\caption{Hysteresis loops recorded at 5K in the fields between $ \pm $±7T for Mn$ _{3} $GaC$ _{x} $, $x$ = 0.8, 1.0 and 1.05) }
\label{fig:mh}
\end{figure}

In order to develop a deeper understanding of the competing ferromagnetic and antiferromagnetic interactions, magnetization loops as a function of field M(H) were recorded at 5K in $H = \pm$7T range and up to 14T in the temperature range of 140K to 280K. The M(H) loops at 5K are shown in Figure \ref{fig:mh}(a). The carbon deficient compound shows a typical soft ferromagnetic behavior with a saturation moment of 3.6 $\mu _{B} $/f.u. and a coercivity of about 91Oe, confirming a FM phase at 5K. Interestingly, the stoichiometric and the carbon rich compound show a {\em S} shaped loop with no saturation up to 7T  and with the virgin curve lying outside the envelope (Figure 4(b)). This behavior suggests presence of competing FM and AFM interactions with contribution of FM growing at the expense of AFM interactions in presence of magnetic field. It can be clearly seen that in the case of C-excess sample, the virgin magnetization curve is completely outside the main loop as compared to that in C-stoichiometric sample indicating stronger antiferromagnetic interactions in Mn$_3$GaC$_{1.05}$. Surprisingly however, the C-excess sample displays higher values of magnetization than C-stoichiometric sample. Plots of magnetization as a function of temperature indicates AFM ground states in both C-stoichiometric and C-excess samples at 5K. This is also corroborated by extrapolating M(H) curve to meet the {\em y-axis}. In both the samples the M $\approx$ 0 at H = 0 indicating abscence of spontaneous magnetization. Therefore the presence of FM interactions in these compounds is rather unexpected and appears to have been induced by external magnetic field. This field induced ferromagnetism possibly explains higher magnetic moment values at 5K in C-excess sample as compared to that of C-stoichiometric sample.

\begin{figure}[h]
\centering
\includegraphics[width=\columnwidth]{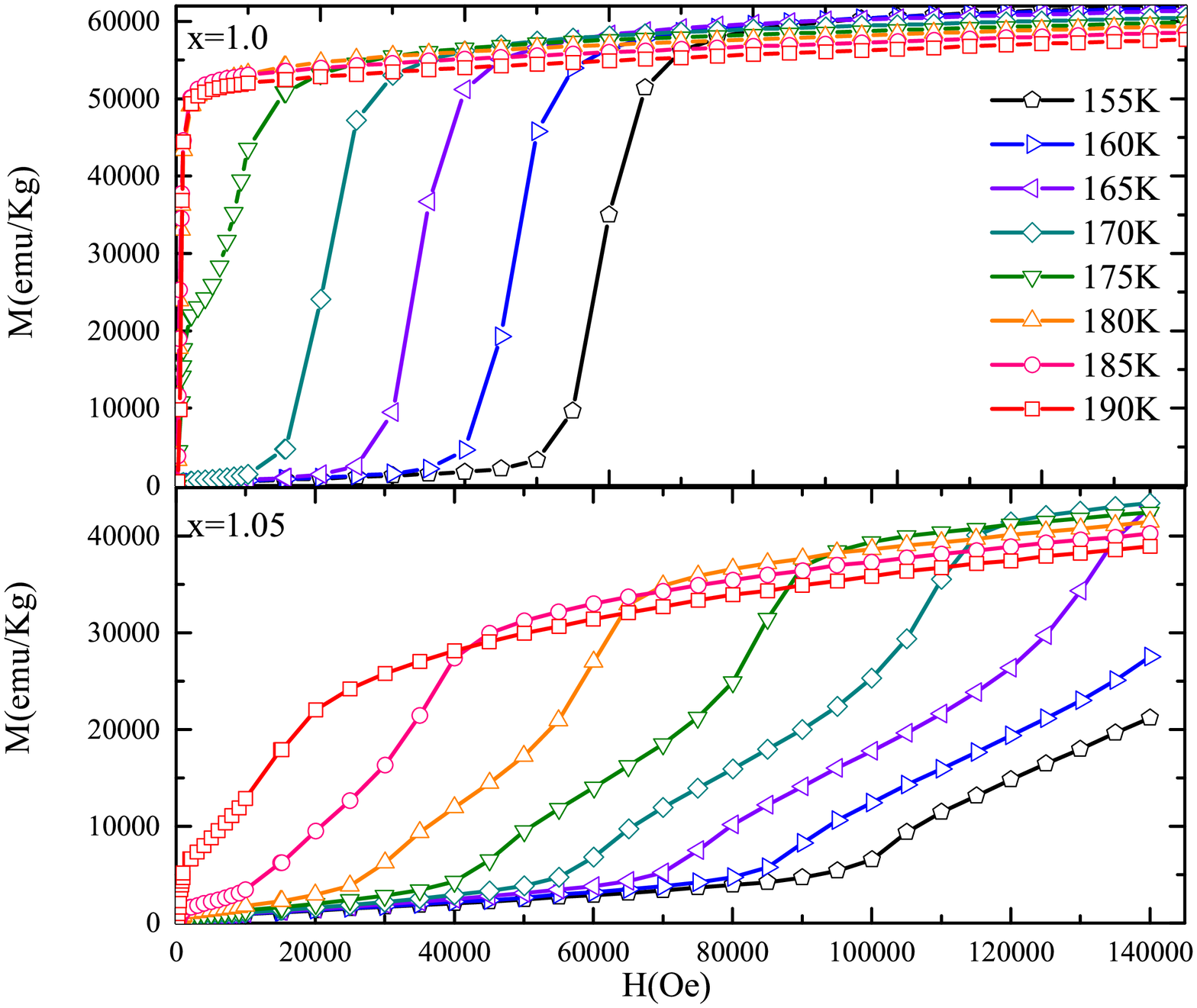}
\caption{Isothermal magnetization curves for Mn$ _{3} $GaC$ _{x} $, $x$ = 1.0 and 1.05 between 155K and 190K
recorded at an interval of 5K in the range 0T $\le H \le$ 14T.}
\label{fig:mh-1st-quad}
\end{figure}

\begin{figure}[h]
\centering
\includegraphics[width=\columnwidth]{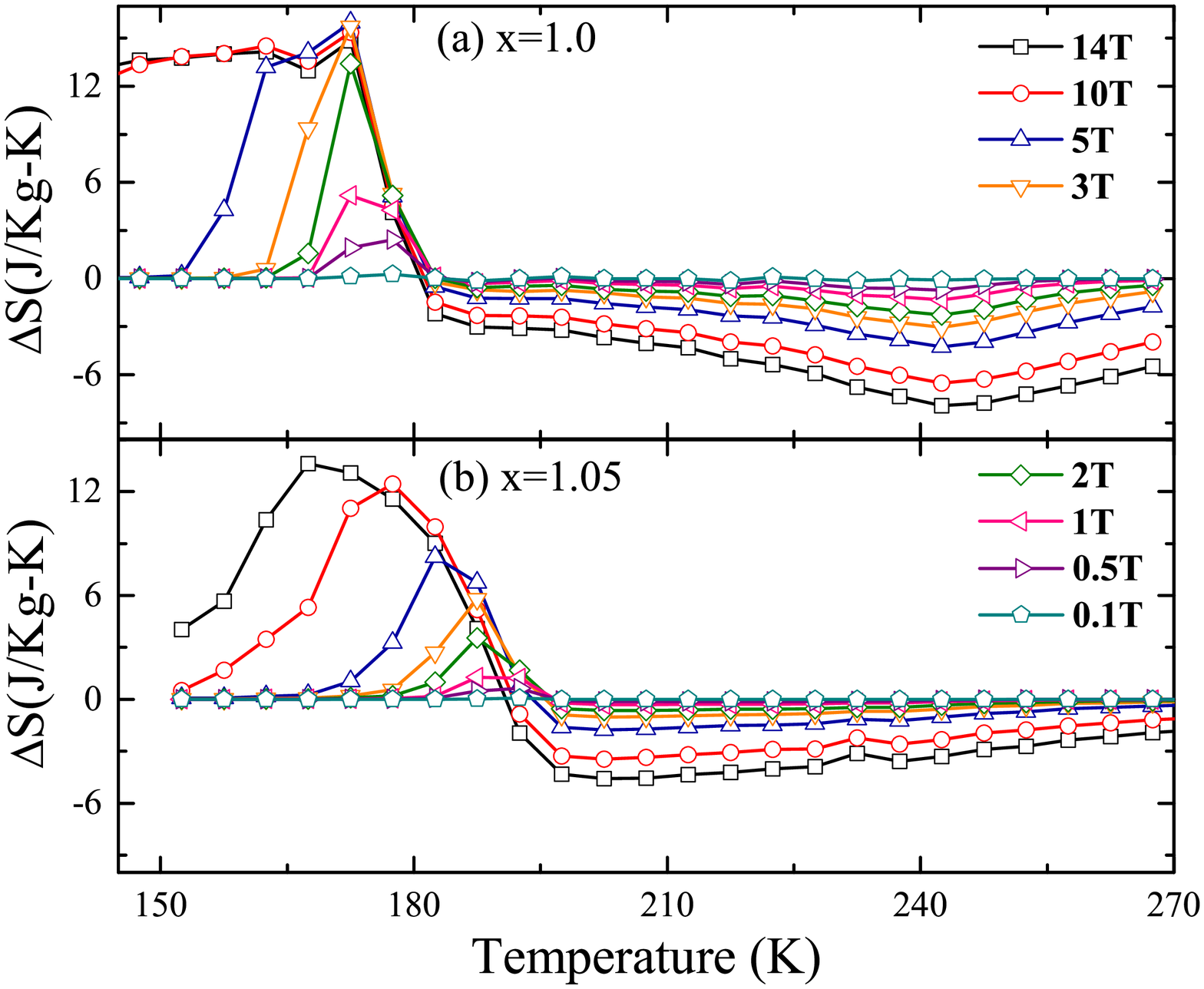}
\caption{Change in entropy ($\Delta S$) calculated from magnetisation isotherms recorded in the temperature range 140K-270K for fields varying from 0.1T to 14T for Mn$ _{3} $GaC$ _{x} $,  $x$ = 1.0 and 1.05. }
\label{fig:mce}
\end{figure}

The presence of field induced ferromagnetism is further confirmed from isothermal magnetization studies conducted in the temperature interval from 140K to 270K.  The isothermal magnetization curves in the limited temperature range of 155K to 190K  and the corresponding entropy plots are presented in Figure \ref{fig:mh-1st-quad} and Figure \ref{fig:mce} respectively. A comparison of isothermal magnetization curves recorded at 190K (ferromagnetic phase) and 155K (antiferromagnetic phase) clearly indicates the field induced effect. In the C-stoichiometric compound, up to H = 5T, the magnetization at 155K is very small, but shows a sharp increase and approaches the ferromagnetic value (M at 190K) for H $>$ 6T. Such a field induced metamagnetic behavior is also seen in the C-excess compound but the transition takes place at much higher fields and over a broad field range. This indicates that even in antiferromagnetic state, ferromagnetism can be induced by external magnetic field. The effect of such a metamagnetic transition can also been seen on the magnetocaloric properties of these compounds. In stoichiometric Mn$_3$GaC an isothermal magnetic entropy change, $\Delta S$ = 16 J/kg-K can be obtained near the first order transition temperature which agrees well the the reported value \cite{e-dias6,e-dias7,e-dias15}. In addition a negative peak in $ \Delta S$ is observed near $T_{C}$ ($\sim$ 240K). This change in entropy reportedly translates to an adiabatic temperature difference of about $\Delta T \approx 4K$ \cite{e-dias8}. Further the $\Delta S$ exhibits very little change as a function of magnetic field (H) for H $\ge$ 2T and remains absolutely constant above H = 6T. This behavior of $\Delta S$ can be directly linked to the metamagnetic transition observed in magnetization plots.

The carbon excess sample also shows a similar positive maximum at its first order transition temperature but with a slightly smaller value of 13.6 J/Kg-K at 14T. There is no negative peak in $\Delta S$ confirming absence of a transition to ferromagnetic state. Furthermore, $\Delta S$  exhibits clear field dependency with its value increasing continuously as well as moves to a lower temperature with increase in $H$. An estimate of $\Delta T$ made using the Clausius Clayperon equation, in low values of magnetic field (H $\sim$ 2T) indicates it to be nearly the same as in case of C-stoichiometric sample. However, the induced ferromagnetism in C-stoichiometric sample causes it to decrease very sharply for higher magnetic field values. This decrease is much slower in the sample which has carbon content in excess.


\section{Conclusions}
In summary, the carbon content critically controls the magnetostructural properties of Mn$_3$GaC type antiperovskites. It is seen that an increase in C content strengthens the antiferromagnetic interactions while weakening the feromagnetic interactions in the compounds. Incorporation of carbon into the lattice increases the unit cell volume which increases the Mn-Mn bond distance and thus weakening the ferromagnetic interactions. The change in character of resistivity from metallic to semiconducting type indicates localization of itinerant Mn $3d$ conduction electrons. This leads to strengthening of Mn-C-Mn antiferromagnetic superexchange interactions.  Furthermore, even in the antiferromagnetic state, magnetic field can be used to tune the magnetic interactions. In Mn$_3$GaC, a applied magnetic field of 6T converts antiferromagnetic state to ferromagnetic one.  Such a conversion cannot be fully achieved in Mn$_3$GaC$_{1.05}$ even up to a field of 14T. This field induced ferromagnetism affects the magnetocaloric properties of the two compounds.

\section*{Acknowledgments}
Authors thank Board of Research in Nuclear Sciences (BRNS) for the financial support under the project 2011/37P/06. M/s Devendra D. Buddhikot and Ganesh Jangam are acknowledged for the experimental assistance. Thanks are also due to Department of Science and Technology, India for the financial support and Saha Institute of Nuclear Physics, India for facilitating the experiments at the Indian Beamline, Photon Factory, KEK, Japan.

\end{document}